\documentclass[aps,jpc,twocolumn,superscriptaddress,10pt,showkeys]{revtex4-1}

\usepackage{graphicx,xcolor}
\usepackage{trfsigns,amsmath,amssymb}
\usepackage{enumitem}
\graphicspath{{./figs/}}	
\begin{document}


\title{Attractor metadynamics in terms of target points in slow--fast systems:\\
       adiabatic vs.\ symmetry protected flow in a recurrent neural network}

\author{Hendrik Wernecke}
\affiliation{Institute for Theoretical Physics, Goethe University Frankfurt, Germany}
\author{Bulcs\'{u} S\'{a}ndor}
\affiliation{Department of Physics, Babe\textcommabelow{s}-Bolyai University, Cluj--Napoca, Romania}
\author{Claudius Gros}
\email[]{gros@th.physik.uni-frankfurt.de}
\affiliation{Institute for Theoretical Physics, Goethe University Frankfurt, Germany}

\date{\today}

\begin{abstract}
In dynamical systems with distinct time scales the time evolution
in phase space may be influenced strongly by the fixed points of 
the fast subsystem. Orbits then typically follow these points, 
performing in addition rapid transitions between distinct branches 
on the time scale of the fast variables. As the branches guide 
the dynamics of a system along the manifold of former fixed points, 
they are considered transiently attracting states and the
intermittent transitions between branches correspond to state
switching within transient--state dynamics.
 
A full characterization of the set of former fixed points, 
the critical manifold, tends to be difficult in high--dimensional
dynamical systems such as large neural networks. Here we point
out that an easily computable subset of the critical manifold, 
the set of target points, can be used as a reference for
the investigation of high--dimensional slow--fast systems. The set
of target points corresponds in this context to the adiabatic 
projection of a given orbit to the critical manifold. Applying
our framework to a simple recurrent neural network, we find that the 
scaling relation of the Euclidean distance between the trajectory 
and its target points with the control parameter of the slow time scale
allows to distinguish an adiabatic regime 
from a state that is effectively independent from target points.
\end{abstract}


\maketitle

\section{Introduction}

Coexisting fixed point attractors such as place 
cells~\cite{moser2008place} are commonly assumed to 
strongly influence cognitive processing in the brain, 
either alone~\cite{hopfield1982neural} or in conjunction 
with feed--forward processing, with the latter being the
case for the episodic memory~\cite{wills2005attractor}.
A system characterized by a single fixed point attractor could
however not be functional on its own, as it would depend
on additional mechanisms to reset the dynamics.
It is hence interesting, that neural activity 
characterized by transitions between multiple meta--stable 
attractors~\cite{rabinovich2008transient} has been discovered 
in the olfactory system of zebrafish~\cite{niessing2010olfactory} 
and in the gustatory cortex~\cite{miller2010stochastic}.
Similar transient state dynamics~\cite{gros2007neural} is also found
in resting state networks in low--frequency contributions
of human fMRI data~\cite{deco2012ongoing}, where it enables 
processes that are associated with cognitive tasks even in the 
complete absence of external stimuli~\cite{deco2011emerging}.
Resting--state brain networks also show complex spatio--temporal 
dynamics, in terms of transitions between states characterized 
by high and low functional connectivities, which resemble transiently 
existing attractor structures \cite{zalesky2014time}. Such 
state--dependent fluctuations may play an important role in 
task--related brain computations~\cite{ritter2015editorial}
such as the interaction of motion and sensation (cf.~\cite{martin2016closed}).

Dynamics involving switching transitions between transiently stable
states has been addressed in the contexts of semantic learning in 
autonomously active networks~\cite{gros2009semantic,gros2009cognitive}, 
within reservoir computing~\cite{lukovsevivcius2009reservoir} and in
networks dominated by heteroclinic orbits~\cite{neves2012computation}.
In case of the latter, periodic orbits are formed when the dynamics
follows heteroclinic connections between saddle points encoding
information in the different states, which
however exist only for symmetry--invariant networks. The details
of the internal dynamics are on the other hand more difficult
to analyze for the case of reservoir computing~\cite{jaeger2001echo},
for which large random network layouts are generically used.

It is well known that transiently attracting states are present in slow--fast
systems~\cite{berglund2006noise,kuehn2015multiple,izhikevich2007dynamical},
i.\,e.\ in dynamical systems with distinct time scales. The fixed 
points of the fast subsystem, which are formally destroyed by 
the slower subsystem, form in this case a manifold that appears 
under several distinct names in the literature, such as slow 
manifold~\cite{berglund2006noise}
and critical manifold~\cite{kuehn2015multiple}. This manifold
corresponds to transiently stable states, and it 
has a pronounced influence
on the overall dynamics, which takes place mostly in the close 
vicinity to this manifold.

The switching dynamics of low dimensional slow--fast dynamical 
systems can be addressed analytically by singular perturbation 
theory~\cite{verhulst2007singular,kuehn2015multiple}, which
has been used to investigate a number of relaxation 
oscillators~\cite{Szmolyan2004}, such as the van--der--Pol 
oscillator and the FitzHugh--Nagumo model for the
dynamics of a single
neuron~\cite{gros2015complex,izhikevich2007dynamical,rocsoreanu2012fitzhugh}.
Further applications of singular perturbation theory
include the study of small Hopfield--type neural networks 
\cite{zheng2014slow}, the analysis of the motion on the  
critical manifold~\cite{verhulst2006dynamics}, of the detailed 
influence of different time scales~\cite{kuehn2012time},
and of the scaling behavior close to 
bifurcations~\cite{kuehn2008scaling, Berglund1999}.

It is generally accepted that a complete hierarchy of time 
scales~\cite{kiebel2008hierarchy} is necessary to describe
the different tasks and activity levels in the brain. Examples
are working memory, short--term and long--term memory and
a range of distinct types of neural plasticity~\cite{tetzlaff2012time}.
Functional hierarchy in recurrent neural networks for motion 
tasks~\cite{yamashita2008emergence} is another phenomenon 
emerging from a separation of time scales.

\subsection{Adiabatic fixed points}

We will start this study by considering attractor
ruins~\cite{linkerhand2013generating} in slow--fast systems,
i.\,e.\ systems that compose of a fast subsystem and an additional slow subsystem,
both being coupled.
Attractor ruins are the remnants of points that are fixed points in the fast
subsystem, but stop being fixed points
(i.\,e., they get ``destroyed'' or ``ruined'') by the interaction with the 
additional slow process (or by noise in noisy setups).

Such states can be found in literature under 
various names, such as attractor
relics~\cite{linkerhand2013generating,gros2014attractor},
transiently attracting states~\cite{gros2014attractor} and
ghosts~\cite{sussillo2013opening}.
Attractor ruins are to be found on the critical manifold and
are hence transiently attracting, but not asymptotically 
stable. They will strongly influence the dynamics of a system 
whenever the coupling to the perturbation is weak.
 
Attractor ruins are defined on the level of the full system,
being however embedded in the phase space of the fast (decoupled)
variables. Hence, as it has been recognized 
previously~\cite{linkerhand2013generating,gros2014attractor},
the evolution of attractor ruins can be cast in terms
of a metadynamics of attractors, which is in turn important
for the modeling of cognitive processes like memory or decision making.

The aim of this study is to investigate the effect of attractor ruins
on the dynamics in the overall system. Therefore we present an efficient
measure to determine the influence of the attractor ruins and distinguish
different dynamical regimes by that.

Furthermore, we introduce here the generic term \textit{adiabatic 
fixed point (AFP)}, denoting all fixed points
of the fast system when the slow variables are fixed but otherwise 
arbitrary. All AFP form, by 
definition, the critical manifold. We note, however, that there are 
stable and unstable AFP.

\subsection{A reference manifold on the critical manifold}

Most of the stable adiabatic fixed points have in general
nothing to do with the long--term behavior of the system. 
We will define in this context the unique set of
\textit{target points} as a subset of AFP corresponding 
to a given trajectory.
We shall concentrate on trajectories that evolve on an
attractor of the full system, i.\,e., on periodic
or chaotic orbits of the full system that are
reached after prolonged times.
For an arbitrary locus on such a trajectory, a target 
point will be defined as the corresponding point 
that would be approached under the time
evolution with fixed slow variables.

As the shape of the manifold of target points only depends 
on the fast subsystem and the trajectory it corresponds to,
this manifold can be used as a reference manifold for the 
motion of the full system. Since slow variables in real world 
applications are not infinitesimally slow, the distance between 
the trajectory and the respective target points,
being larger than zero, is an important 
measure for the assessment of the influence a target point and
hence the critical manifold as a whole exerts on the dynamics. 
The critical manifold is both uniquely defined and 
numerically accessible; the manifold of target points, being a subset of
the critical manifold, is also uniquely defined, as it corresponds to 
a certain trajectory. Therefore we will use the manifold of target points
in this approach as a low dimensional reference manifold to characterize
the corresponding dynamics in a potentially high dimensional phase space.

\subsection{Adiabatic and non--adiabatic regimes}

Analyzing the distance between a given orbit and its associated
target points we discriminate two qualitatively different
dynamical regimes:
\begin{itemize}
\item In the \textit{adiabatic regime} the dynamics is effectively
 influenced by target points, with the average distance of the trajectory
 to the target points scaling with the ratio of the slow
 and the fast time scale.
 We emphasize that the dynamics of the slow subsystem does not vanish in
 the adiabatic regime. The term relates rather to a regime of parameters,
 for which the trajectory follows the target points (which are \textit{adiabatic}
 fixed points) closely, interseeded with jumps between different branches of
 the critical manifold.
 The corresponding different branches of the manifold of target points 
 represent different transiently attracting states.
\item In the \textit{non--adiabatic regime} the dynamics effectively
 decouples from the target points, with the distance to the target 
 points being in essence invariant with respect to changes of the relative 
 time scale.
\end{itemize}
One may discriminate equivalently between regimes in which a
perturbation expansion in the slow time scale converges 
or diverges, respectively~\cite{kuehn2015multiple}.
Our approach aims on the other hand to develop tools
suitable for the numerical study of high--dimensional 
slow--fast systems. We find that the study of target points,
which are straightforward to determine numerically, allows 
in this context to distinguish between the adiabatic and
the non--adiabatic regime.

\begin{figure*}[t]\centering
\includegraphics[width=0.99\textwidth]{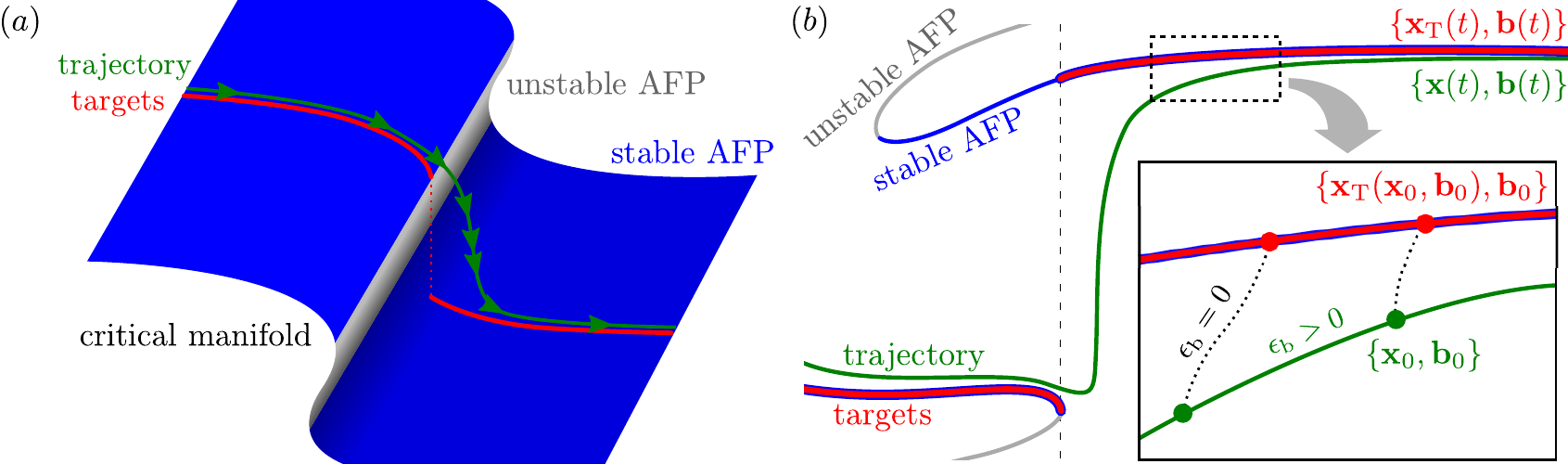}
\caption{\label{fig:targetsketch}
Sketch of the effect of target points on a trajectory. ($a$)
The target manifold (red line) is a subset of the manifold of 
all stable fixed points in the fast subsystem (blue sheets).
Note that the critical manifold consists of stable AFP (blue surfaces) and
unstable AFP (grey surface).
The target points correspond uniquely to a given trajectory 
(green line), which follows the target points slightly delayed.
($b$)
At every point $\{\mathbf{x}_0,\mathbf{b}_0\}$ on the trajectory 
(green bullets in the inset) the fast subsystem $\mathbf{x}(t)$ is attracted 
by the uniquely corresponding target point 
$\{\mathbf{x}_\text{T}(\mathbf{x}_0,\mathbf{b}_0),\mathbf{b}_0\}$ 
(red bullets). For $\epsilon_\text{b}=0$ the 
system would converge (dashed lines) to the target point. For a
finite flow in the slow subsystem $\mathbf{b}(t)$, viz when 
$\epsilon_\text{b}>0$, the system evolves, on the other side, along
the trajectory (green line). The set of target points corresponds 
to a map of the trajectory $\{\mathbf{x}(t),\mathbf{b}(t)\}$ 
to the critical manifold.
}
\end{figure*}

\subsection{Three--site recurrent neural networks}

In the second part of this study we apply our framework to a network 
of continuous--time rate--encoding neurons, which has been
shown previously to exhibit non--trivial dynamical states~\cite{markovic2012intrinsic}.
As we are mainly interested in how AFP influence the intrinsic behavior of the
network, we restrict ourselves to autonomously active networks,
i.\,e.\ networks without external input. A separation between the 
time scale of the membrane potential and an intrinsic neural parameter, 
the threshold, is present in this system.

For concreteness we study a three--neuron network allowing for
an in--depth understanding of the resulting dynamical states. 
Computing the transiently attracting states
(as connected sets of target points) and the distance 
measures quantifying the influence of target points on 
the actual dynamics, leads to the characterization 
of two distinct dynamical regimes, i.\,e.\ the adiabatic and the 
non--adiabatic regime. We conclude with an analysis of the 
transition between these two regimes and of the role
of target points for chaotic motion.

\section{Theory\,---\,Attractor metadynamics in slow--fast dynamical systems}

We consider with
\begin{equation}
\begin{array}{rcl}
 \dot{x}_i &= &f_i(\mathbf{x},\mathbf{b}) \\
 \dot{b}_j &= &\epsilon_\text{b} g_j(\mathbf{x},\mathbf{b})
\end{array}\label{eq:dynsys}
\end{equation}
$N+M$ dimensional slow--fast dynamical systems, where we have
denoted with $\mathbf{x}=\mathbf{x}(t)=\{x_i(t)\}_{1\leq i\leq N}$ 
the fast variables and with 
$\mathbf{b}=\mathbf{b}(t)=\{b_j(t)\}_{1\leq j\leq M}$ the slow variables.
The slow time scale is set by $\epsilon_\text{b}$, which is also the 
ratio of time scales. 

The set of fixed points in the fast subsystem, i.\,e.\
$\mathrm d\mathbf{x}/\mathrm dt=0$ for given and fixed $\mathbf{b}$,
is of special interest for the analysis of slow--fast dynamical 
systems. The entirety of these points, which are the intersections
of the nullclines $\mathrm dx_i/\mathrm dt=0$ of the fast 
subsystem~\cite{simonyi1967dynamics}, are generally termed
slow manifold~\cite{berglund2006noise} or critical 
manifold~\cite{kuehn2015multiple}.
Please note that the term slow manifold is used in different 
contexts and fields and also with different definitions as 
already pointed out by Lorenz~\cite{lorenz1992slow}
(cf.~\cite{berglund2006noise,kuehn2015multiple,verhulst2007singular}).
In the context of this work we will refer to the set of fixed points
in the fast subsystem as critical manifold, as this nomenclature is widely used
in the field.
Here we constrain ourselves to a fast subsystem with only fixed point 
attractors. This is for instance the case if the fast subsystem is a gradient system,
e.\,g.\ when the fast dynamics is derived from a generating
functional~\cite{markovic2012intrinsic,markovic2010self}. 
The different branches of the critical
manifold are therefore exclusively composed of isolated fixed points. They are
often well characterized, for real--world systems, in terms of
their physical and/or neurobiological properties~\cite{deco2012ongoing}.

\subsection{Adiabatic fixed points (AFP) and target points}

In physics terminology the limit $\epsilon_\text{b}\to0$,
i.\,e.\ when the slow subsystem is infinitely slow, is termed the adiabatic limit,
such as in the Born--Oppenheimer approximation~\cite{born1927quantentheorie}, 
where the slow movement of the atomic nuclei can be treated as parametric 
variables when addressing the relatively fast dynamics of the electrons.
In the case $\epsilon_\text{b}=0$ the configuration~$\mathbf{b}$ of the slow
subsystem is constant and can be treated as parameters.
We will hence use the term adiabatic fixed point (AFP) for the fixed points of 
the fast subsystem.
Note that the set of all AFP is equivalent to the critical
manifold~\cite{kuehn2015multiple} of the system.
 
Considering a generic state $\{\mathbf{x}_0, \mathbf{b}_0\}$
as the starting point of a solution
$\mathbf{x}(t')\lvert_{\mathbf{b}=\mathbf{b}_0}$
of Eq.~(\ref{eq:dynsys}) for fixed slow variables~
$\mathbf{b}=\mathbf{b}_0$, we define with
\begin{equation}
\mathbf{x}_\text{T}(\mathbf{x}_0,\mathbf{b}_0)=
\lim_{t'\to\infty}\mathbf{x}(t')\rvert_{\mathbf{b}=\mathbf{b}_0}
\label{eq:xT0}
\end{equation}
the fast component of the target point 
$\{\mathbf{x}_\text{T}(\mathbf{x}_0,\mathbf{b}_0),\mathbf{b}_0\}$
corresponding to the respective starting point. Here we take 
$\mathbf{x}_0=\mathbf{x}(t_0)$ and $\mathbf{b}_0=\mathbf{b}(t_0)$ 
to be the fast and respectively slow components of a point on 
a trajectory $\{\mathbf{x}(t),\mathbf{b}(t)\}$ in the full 
system, as parameterized by the time $t=t_0$. The mapping 
(\ref{eq:xT0}) is unique for a given pair $(\mathbf{x}_0,\mathbf{b}_0)$,
depending furthermore not on $\epsilon_\text{b}$, which is the ratio
of the time scales.
All target points are AFP and a set of target points corresponding to
a given trajectory is a subset of the critical manifold.
This set of target points can thus be used as a reference manifold that allows
to analyze the dynamics of the overall system relative to the critical manifold.
 
The relation of target points, trajectory and critical manifold
is sketched schematically in Fig.~\ref{fig:targetsketch}~($b$). 
We also present in Table~\ref{tab:computing} 
(cf.~Appendix~\ref{sec:app_code}) a pseudo--code 
for the numerical computation of AFP and 
target points.

\subsection{Kinetic energy of phase space evolution}

Target points can be computed via a straightforward
evolution of the equations of motion of the fast subsystem
for fixed slow variables $\mathbf{b}$. Unstable adiabatic 
fixed points may be found~\cite{sussillo2013opening}, in 
addition, by minimizing the kinetic energy
\begin{equation}
q_x = \frac{\dot{\mathbf{x}}^2}{2}=\frac{1}{2}\sum\limits_{i=1}^N\dot{x}_i^2 
\label{eq:q_flow}
\end{equation}
of the flow in the fast subsystem. A minimum with $q_x=0$ 
corresponds to a fixed point of the fast subsystem, regardless 
of its stability. This is used to compute AFP numerically 
(cf.~Appendix~\ref{sec:app_code}).

The zeros of $q_x$ are generically not minima of the 
kinetic energy $q=(\dot{\mathbf{x}}^2+\dot{\mathbf{b}}^2)/2$ 
of the full system, for which also the flow in the slow 
subsystem, $\dot{\mathbf{b}}$, needs to vanish. The same holds
for slow points, which are defined by a vanishing $q_x$ and
small values of $q$, see~\cite{sussillo2013opening}.

\subsection{Attractor metadynamics}

The set of target points $\{\mathbf{x}_\text{T}(t),\mathbf{b}(t)\}$,
where
\begin{equation}
\mathbf{x}_\text{T}(t) \equiv \mathbf{x}_\text{T}(\mathbf{x}(t),\mathbf{b}(t))
\label{eq:xTt}
\end{equation}
corresponds to the mapping~(\ref{eq:xT0})
of a trajectory $\{\mathbf{x}(t),\mathbf{b}(t)\}$ onto the critical manifold.
As mentioned before,
we shall only consider trajectories on attractors for this study.
The set of target points hence has the same dimensionality
as the respective attractor, e.\,g.\ one dimensional for a limit cycle
or fractal for a chaotic attractor (cf.~Sect.~\ref{sec:3neurons}).
In typical slow--fast systems the embedding dimension of the critical manifold
is usually much lower than the dimension of the full system. The dimension
of a set of target points corresponding to a given trajectory, being a subset
of the critical manifold, is thus even lower, reducing therefore the complexity
of the analysis.

One defines as attractor metadynamics the time evolution
of the target points Eq.~(\ref{eq:xTt}), which can be 
continuous or characterized by jumps between distinct
sets of target points.
Distinct branches of the critical manifold can often be 
classified in real world applications, e.\,g.\ in the 
neurosciences when using a slow feature 
analysis~\cite{franzius2011invariant}.
States on the same branch of
the critical manifold are then lumped together, with distinct
branches corresponding to different objects, such as
`chair' and `table'. Target points continuously connected,
compare Fig.~\ref{fig:targetsketch}, can then be
considered a set of states corresponding to the same
object and hence as a transiently attracting state.
The mapping (\ref{eq:xTt}) therefore implies that the
flow moves from one transiently attracting state to another 
whenever the respective target point jumps from one branch 
of the critical manifold to the next.
This classification is considered important
especially in the neurosciences, where attracting states 
guide decision making, memory storage and 
recognition~\cite{rabinovich2008transient}.

\subsection{Distance to target points}\label{sec:dist}

As a measure of the influence a given target point exerts
on the trajectory we will consider the Euclidean 
distance~$d(t)$,
\begin{equation}
d(t)=\sqrt{ (\mathbf{x}(t)-\mathbf{x}_\text{T}(t))^2}\,,
\qquad
\langle d\rangle =\frac{1}{\tau}\int\limits_0^\tau\!\mathrm dt\;d(t)
\label{eq:dist}
\end{equation}
between the fast components of the trajectory 
$\{\mathbf{x}(t),\mathbf{b}(t)\}$ and the corresponding
target points $\{\mathbf{x}_\text{T}(t),\mathbf{b}(t)\}$,
see Eq.~(\ref{eq:xTt}). The corresponding
time--average $\langle d\rangle$ has been defined in (\ref{eq:dist})
for the case of a limit cycle with period $\tau$, with
a corresponding straightforward generalization for
chaotic attractors.

The average distance $\langle d\rangle$ vanishes, per
construction, in the adiabatic limit $\epsilon_\text{b}\to0$.
For finite $\epsilon_\text{b}>0$ we find, on the other hand,
two regimes, with $\langle d\rangle$ scaling with
$\epsilon_\text{b}$ in the \textit{adiabatic regime}, but
not in the \textit{non--adiabatic regime}.

Apart from the average distance $\langle d\rangle\to0$,
one can also evaluate the probability density function $\rho(d)$
of the distances over a given attractor. The corresponding 
cumulative distribution function of distances
\begin{equation}
P(d)=\int\limits_{0}^{d}\!\!\mathrm du\;\rho(u)
\label{eq:cumdist}
\end{equation}
can then be used to characterize regimes.
As a trajectory stays close to the target points for prolonged time spans in the
adiabatic regime one finds contributions at small distances
$d\sim\epsilon_\text{b}\ll1$ for the cumulative distribution.
A second contribution comes from the jumps between the different branches of
target points at larger distances $d\gg\epsilon_\text{b}$ (usually $d\sim1$).
However, in the non--adiabatic case only the latter contribution, at larger
distances, exists, as the system comes close to the target points only occasionally.
A lack of contributions at small distances $d\sim\epsilon_\text{b}$
is an evidence that the dynamics is in the non--adiabatic regime.

\begin{figure}[t]\centering
\includegraphics[width=0.40\textwidth]{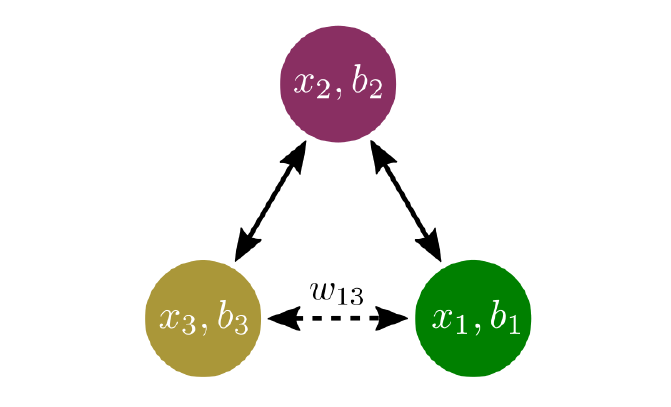}
\caption{\label{fig:sketch}
Scheme of the three--neuron system. The second neuron 
(top) is exciting the other two neurons 
(solid arrows) and vice versa. The first and the third 
neuron (bottom) are coupled by an inhibitory
connection $w_{13}<0$ (dashed arrow).
}
\end{figure}

\section{Results}

\subsection{Three--neuron system}\label{sec:3neurons}

To investigate the effect of target points on the overall dynamics,
we consider for this study a small neural network (cf.~Fig.~\ref{fig:sketch})
of three rate--encoding continuous--time point neurons~\cite{markovic2012intrinsic}.
Networks of three neurons have already served as model systems in different
contexts such as modeling pacemaker circuits~\cite{prinz2004similar},
the stomatogastric ganglion in lobster~\cite{miller1982mechanisms} or
neural motifs~\cite{morgan2008nonrandom}.
Here we chose the three--neuron layout to study the fundamental properties 
of AFP and target points, as it allows  for a full investigation
of its non--trivial dynamics and phase diagram.

The fast subsystem corresponds in this case to the time evolution
\begin{equation}
\dot{x}_i=-x_i+\sum\limits_{j=1}^{N}w_{ij}y_j\;,
\label{dot_x}
\end{equation}
of the membrane potential $x_i$, with $w_{ij}>0$ and
$w_{ij}<0$ denoting excitatory and inhibitory connection $j\to i$
respectively. The corresponding firing rate
$y_i\in[0,1]$, the neural activity, is a sigmoidal
function of the respective membrane potential:
\begin{eqnarray}
y_i=\frac{1}{1+\exp\left(a_i(b_i-x_i)\right)}\,.
\label{eq:sigmoidal}
\end{eqnarray}
We have used constant gains $a_i\equiv a=6$, whereas
the threshold $b_i=b_i(t)$ is adapting slowly~\cite{triesch2005gradient}
on extended time scales $1/\epsilon_\text{b}\gg1$,
\begin{equation}
\dot{b}_i=  \epsilon_\text{b}\, 2a(y_i-1/2)~.
\label{eq:dot_b}
\end{equation}
In the context of this model the ratio of times scales $\epsilon_\text{b}$
is also called the adaption rate of the slow variables $b_j$.
The time evolution Eq.~(\ref{eq:dot_b}) of the slow variables attempts to drive
the dynamics towards $y_i\to1/2$ as $\dot{b}\to0$, which is the fixed point of
the full system (cf.~Appendix~\ref{sec:app_fp}).

The state of a neuron is hence described by the tuple 
$\{x_i(t), b_i(t)\}$. For the particular three--neuron system 
sketched in Fig.~\ref{fig:sketch}, the dynamics of the fast 
variables $x_i$ is given by~\cite{linkerhand2013generating} 
\begin{equation}
\begin{array}{rcl}
 \dot{x}_1&=&-x_1+y_2+w_{13}y_3 \\
 \dot{x}_2&=&-x_2+y_1+\phantom{w_{23}}y_3\\
 \dot{x}_3&=&-x_3+y_2+w_{13}y_1\,,
\end{array}
\label{eq:dot_x}
\end{equation}
with $w_{13}=w_{31}<0$ being inhibitory.
All remaining synaptic weights are unity $w_{12}=w_{21}=w_{23}=w_{32}=1$
and thus excitatory connections.
For the results presented here we computed the numerical solution 
of the ODE system Eqs.~(\ref{eq:dot_b}) and~(\ref{eq:dot_x}) 
performing a fourth order Runge--Kutta integration algorithm with Fehlberg
tableau~\cite{fehlberg1969low} 
using step size $\mathrm dt=10^{-2}$.
For computing the AFP we used a minimization algorithm with a BFGS
strategy~\cite{broyden1970convergence,shanno1970conditioning} provided by
the \texttt{dlib} optimization library~\cite{dlib} for the \texttt{C++}
programming language.

\subsubsection{Symmetries}\label{sec:sym}

The network shown in Fig.~\ref{fig:sketch} is symmetric under 
the exchange $1\leftrightarrow3$ of the first and the third 
neuron, a reflection symmetry.
For the special case $w_{13}=-1$ the additional $C3$ rotational
symmetry $\mathbf{x}\to\tilde{\mathbf{x}}$
and $\mathbf{b}\to\tilde{\mathbf{b}}$, with
\begin{equation}
\begin{alignedat}{3}
\tilde{x}_1&=-x_3,\qquad&\tilde{x}_2&=x_1+1,\qquad&\tilde{x}_3&=x_2-1\\
\tilde{b}_1&=-b_3,&\tilde{b}_2&=b_1+1,&\tilde{b}_3&=b_2-1
\label{eq_C3}
\end{alignedat}
\end{equation}
is present. Eq.~(\ref{eq_C3}) can be verified by inspecting
Eqs.~(\ref{eq:dot_b}) and (\ref{eq:dot_x}). This leads to
\begin{equation}
\tilde{y}_1=1-y_3,\quad \tilde{y}_2=y_1, \quad \tilde{y}_3=y_2\,.
\end{equation}
Applying this iteration three times yields the identity transformation,
Eq.~(\ref{eq_C3}) is hence equivalent to a $C3$ symmetry.

\begin{figure}[]\centering
\includegraphics[width=0.40\textwidth]{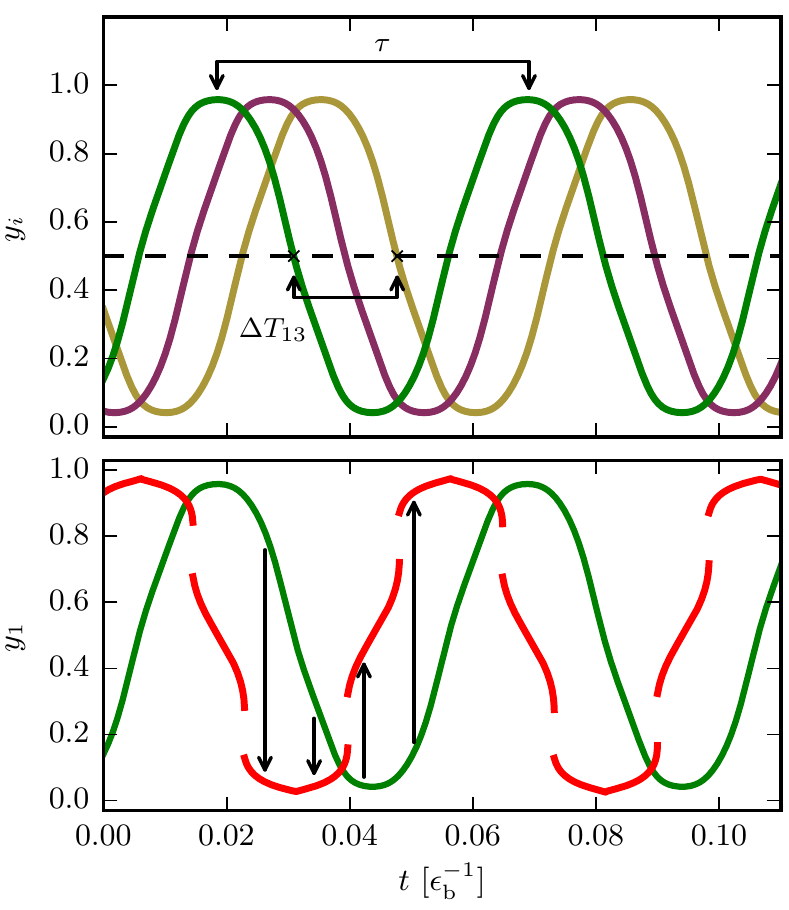}
\caption{\label{fig:traj_nonad}
Firing rates  of the three neurons for $\epsilon_\text{b}=8\cdot10^{-4}$
in the $C3$ symmetric case $w_{13}=-1$ 
(indicated by the magenta bullet in Fig.~\ref{fig:phaseshift}) 
over rescaled time.
(\textit{top}) Green/violet/yellow denote $y_1$/$y_2$/$y_3$;
the relative phase shift $\varDelta_{13}=\varDelta T_{13}/\tau=1/3$
is the ratio of the time shift $\varDelta T_{13}$ between two
consecutive Poincar\'{e} sections $y_i=1/2$, $\dot{y}_i<0$ of 
the first and third neuron, and the oscillation period $\tau$.
(\textit{bottom})~$y_1$--component of the target points (red)
and of the trajectory (green), with the arrows indicating the 
behavior of the system in the adiabatic limit 
$\epsilon_\text{b}\to0$ (cf. Fig.~\ref{fig:targetsketch}).
}
\end{figure}

\subsubsection{Dynamics of the three--neuron system}
Before analyzing the transiently attracting states of the 
three--neuron system we briefly discuss here
the possible types of long--term dynamics in the network,
i.\,e.\ attractors in the phase space of the full system,
for different cases of 
the inhibitory weight~$w_{13}$ and of the time scale 
difference~$\epsilon_\text{b}$.

\paragraph{$C3$ symmetry}
Due to the $C3$ symmetry of the three--neuron system for 
$w_{13}=-1$ discussed in Sect.~\ref{sec:sym},
we find a traveling wave solution where all neurons show 
the same activities $x_i(t)=x(t-\theta_i)$ with period $\tau$, 
albeit with distinct phases $\theta_i$, shifted respectively by $\tau/6$.
One can easily prove that this type of motion is always an exact 
solution in the case of $w_{13}=-1$ (cf. Appendix~\ref{sec:app_tw}).

In the top panel of Fig.~\ref{fig:traj_nonad} we show an example of 
the traveling waves solution for $w_{13}=-1$, $\epsilon_\text{b}=8\cdot10^{-4}$
with green/violet/yellow encoding $y_1/y_2/y_3$. Time has
been rescaled with respect to $1/\epsilon_\text{b}$.  
For a further quantification we define the dimensionless 
relative phase shift
\begin{align}
\varDelta_{13}&=\varDelta T_{13}/\tau
\label{Delta_13}
\end{align}
between the first
and the third neuron, where $\varDelta T_{13}$ is the time difference
of two consecutive intersections of $y_1$ and $y_3$ with the
Poincar\'{e} plain $y_i=1/2$, $\dot{y}_i<0$ 
(cf. top panel of Fig.~\ref{fig:traj_nonad}).
The traveling waves solution is characterized hence 
by $\varDelta_{13}=1/3$.

The traveling waves motion is an exact solution for
all $\epsilon_\text{b}$, but stable only for
$\epsilon_\text{b}>6\cdot10^{-5}$. For all
$\epsilon_\text{b}>6\cdot10^{-5}$ the motion is
hence symmetry protected \cite{pollmann2012symmetry} in the sense that the
relative phase shift (\ref{Delta_13}) is
independent of~$\epsilon_\text{b}$, viz
$\varDelta_{13}\equiv1/3$.

\begin{figure}[t]
\centering
\vspace{4mm}
\includegraphics[width=0.40\textwidth]{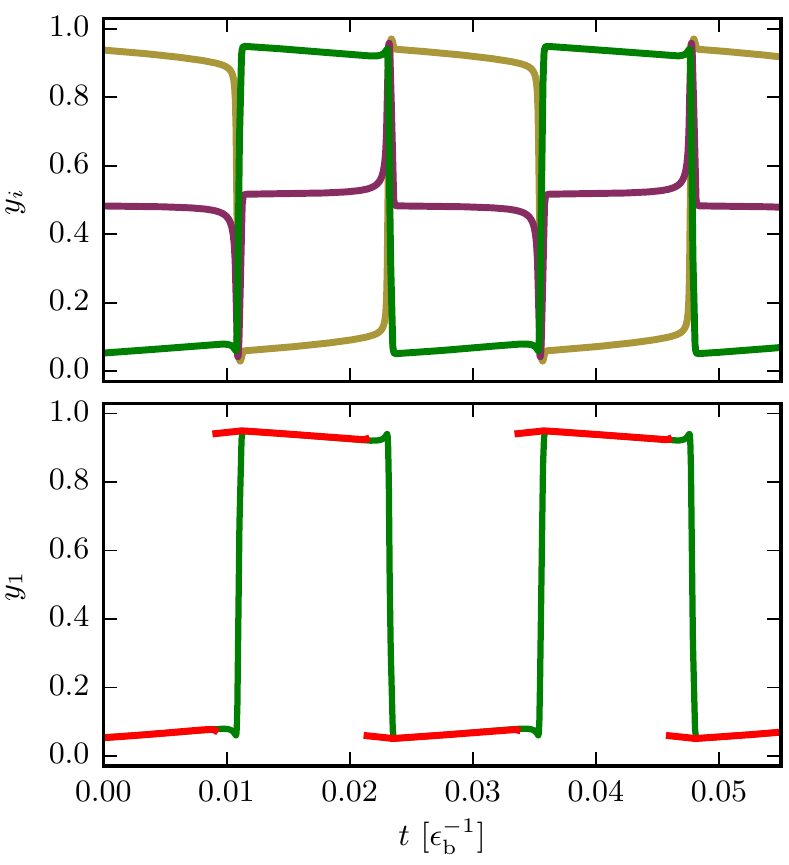}
\caption{\label{fig:traj_ad}
Firing rates  of the three neurons for $\epsilon_\text{b}=10^{-5}$
in the $C3$ symmetric case $w_{13}=-1$ 
(indicated by the cyan square in Fig.~\ref{fig:phaseshift})
over rescaled time.
(\textit{top})~green/violet/yellow denote $y_1$/$y_2$/$y_3$.
(\textit{bottom})~ the $y_1$--component of the target points
(red) and of the trajectory $y_1$ (green).
The system performs an anti--phase flip--flop oscillation
of the first and the third neuron, $\varDelta_{13}=1/2$.
}
\end{figure}

In the top panel of Fig.~\ref{fig:traj_ad} we show 
the trajectory for $\epsilon_\text{b}=10^{-5}$,
viz in the phase where the traveling wave solution
is not anymore stable. The activity levels of
the first and the third neuron now approach mutually either 
full or low activity states, $y_i\approx1$ and $y_i\approx0$
respectively, while the second neuron stays mostly 
at $y_2\approx 1/2$ (half active). The relative phase 
shift is $\varDelta_{13}=1/2$, which is the maximal
possible value.

The system thus flips between the two states
$\mathbf{y}=(y_1,y_2,y_3)\approx(0,1/2,1)$ and
$\mathbf{y}\approx(1,1/2,0)$.
The flipping time of this flip--flop oscillation
scales inversely with the adaption rate~$\epsilon_\text{b}$.

\begin{figure}\centering
\includegraphics[width=0.495\textwidth]{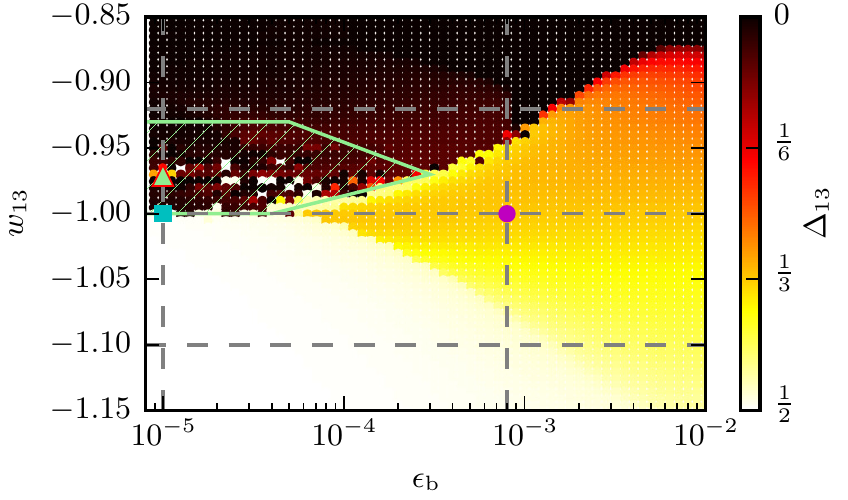}
\caption{\label{fig:phaseshift}
Phase diagram, as a function of the adaption 
rate~$\epsilon_\mathrm{b}$ and of the inhibitory 
weight $w_{13}$, where the color encodes the phase
difference $\varDelta_{13}$ of the first and the third 
firing rate of the neuron, normalized with the period of 
the oscillation (color bar; compare Eq.~(\ref{Delta_13})).
The two main regimes show in--phase oscillations $\varDelta_{13}=0$
of all neurons (dark) or anti--phase oscillation $\varDelta_{13}=1/2$
of neurons 1 and 3 (bright). Close to the symmetric axis 
$w_{13}=-1$ there exists a region in the parameter space with phase 
shifts in between these extreme cases. The dashed gray horizontal lines 
mark the cuts through the parameter section considered
in Figs.~\ref{fig:phasediag_epsb}, \ref{fig:disttoAFP}
and \ref{fig:phasediags_w13_1} (left panel). The
solutions corresponding to the 
magenta bullet ($\epsilon_\text{b}=8\cdot10^{-4}, w_{13}=-1$)
and to the cyan square ($\epsilon_\text{b}=10^{-5}, w_{13}=-1$) 
are shown in Fig.~\ref{fig:traj_nonad} and Fig.~\ref{fig:traj_ad} 
respectively. Disperse patches of chaotic motion can be found in 
the green shaded area. One example is the red framed 
triangle ($\epsilon_\text{b}=10^{-5}, w_{13}=-0.9709$;
compare Fig.~\ref{fig:trajchaos}).
}
\end{figure}

\paragraph{No $C3$ symmetry}
In Fig.~\ref{fig:phaseshift} we outline the distinct
dynamical regimes observed for the general case of
arbitrary inhibitory weights close to the symmetric
case.
The relative  phase shift $\varDelta_{13}$ is color--coded 
and shown as a function of the adaption rate 
$\epsilon_\mathrm{b}$ and of the inhibitory weight 
$w_{13}$.

On the horizontal center axis $w_{13}=-1$, which represents 
the case of a $C3$ symmetric system, the symmetry protected 
traveling wave solution with $\varDelta_{13}=1/3$ occurs
for $\epsilon_\text{b}>6\cdot10^{-5}$ (cf.~Fig.~\ref{fig:traj_nonad})
and the flip--flop oscillation $\varDelta_{13}=1/2$ for 
$\epsilon_\text{b}<6\cdot10^{-5}$
(cf.~Fig.~\ref{fig:traj_ad}).

Far off the symmetric axis only two possible states exist:
either the system performs a flip--flop oscillation 
($\varDelta_{13}=1/2$ and $w_{13}<-1.15$) or an in--phase 
oscillation ($\varDelta_{13}=0$ and $w_{13}>-0.85$).
In the latter case all three neurons fire synchronously and
switch between a state $\mathbf{y}\approx(1,1,1)$ of high activity
and a state $\mathbf{y}\approx(0,0,0)$ of low activity.

\begin{figure*}[th]\centering
\includegraphics[width=0.99\textwidth]{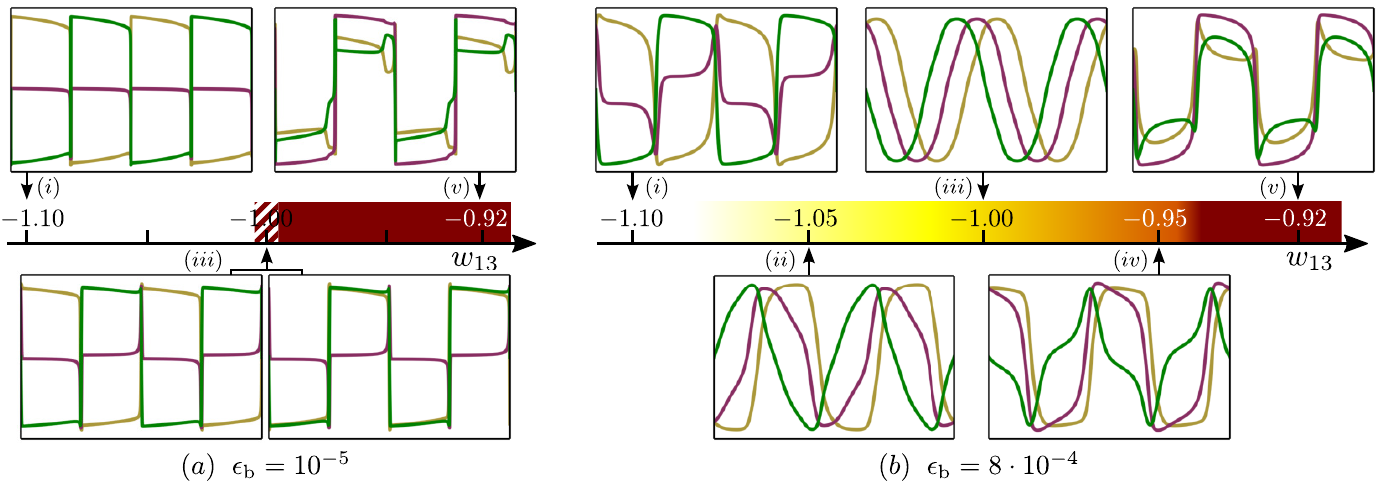}
\caption{\label{fig:phasediags_w13_1}
Vertical cuts through the phase diagram shown in Fig.~\ref{fig:phaseshift}.
The color of the bar encodes the phase shift $\varDelta_{13}$
as a function of $w_{13}$.
The insets show the firing
rates of all three neurons (cf. the upper panels of Figs.~\ref{fig:traj_nonad}
and \ref{fig:traj_ad}).
($a$) For $\epsilon_\text{b}=10^{-5}$ the transition from 
anti--phase to in--phase oscillations occurs near the symmetric 
point~$w_{13}=-1$. The area marked by stripes shows bistability
and both types of attractors coexist.
($b$) For~$\epsilon_\mathrm{b}=8\times10^{-4}$ the
firing rates change in a smooth manner from anti--phase 
oscillation to traveling waves, and with a small jump
to in--phase behavior.
}
\end{figure*}

These two main types of motion follow directly from the topology
of the network illustrated in Fig.~\ref{fig:sketch}.
For relatively strong inhibition $w_{13}<-1$ the first and the 
third neuron suppress each other mutually, they can therefore not 
be active at the same time. A dominance of inhibition thus leads 
to a flip--flop oscillation. Weak inhibition $w_{13}>-1$ allows
on the other side both the first and the third neuron to be 
active at the same time, and therefore also the second neuron,
leading to an in--phase oscillation of all three neurons.

The situation is more complicated in the inner region of
$-1.15<w_{13}<-0.85$ (cf. Fig.~\ref{fig:phaseshift}),
where inhibitory and excitatory weights are similar in 
magnitude. The relative phase shift $\varDelta_{13}$
varies continuously from its maximal value $1/2$ to its minimal value $0$,
when increasing $w_{13}$ for $\epsilon_\text{b}>6\cdot10^{-5}$,
taking the value $\varDelta_{13}=1/3$ right at the symmetry
point $w_{13}=-1$, where the traveling wave solution,
illustrated in Fig.~\ref{fig:traj_nonad} exists.

Below, we will discuss further horizontal and vertical cuts
through parameter space, indicated by the dashed gray
lines in Fig.~\ref{fig:phaseshift}, as well as the occurrence of
chaotic attractors for certain parameter values to be found in the area
that is green shaded in Fig.~\ref{fig:phaseshift}.

In Fig.~\ref{fig:phasediags_w13_1} we present the sketch
of the behavior along the previously mentioned two cuts of the phase diagram
for $\epsilon_\text{b}=10^{-5}$ and $\epsilon_\text{b}=8\cdot10^{-4}$
(cf.~Fig.~\ref{fig:phaseshift}).
For low adaption $\epsilon_\text{b}=10^{-5}$ as shown
in Fig.~\ref{fig:phasediags_w13_1}~($a$)
we only find adiabatic behavior:
either anti--phase oscillations~($i,iii$) or in--phase oscillations~($v$). 
Close to the symmetric transition point $w_{13}=-1$ 
both types of attractors can be found depending on the
initial conditions.
As mentioned before, the anti--phase oscillation corresponds 
to a switching between the two anti--symmetric transiently attracting states
$\mathbf{y}\approx\{0,1/2,1\}$ and  $\mathbf{y}\approx\{1,1/2,0\}$. 
In the case of the in--phase oscillation this is a 
periodic switching between the symmetric transiently attracting states
$\mathbf{y}\approx\{0,0,0\}$ and $\mathbf{y}\approx\{1,1,1\}$.
For low adaption rate we thus find a direct transition between the
two different types of attractors of the adiabatic regime.
With a larger adaption $\epsilon_\text{b}=8\cdot10^{-4}$
in Fig.~\ref{fig:phasediags_w13_1}~($b$) at the transition from
strong inhibition to weak inhibition the traveling waves regime appears.
The panels in Fig.~\ref{fig:phasediags_w13_1}~($b$) depict five
examples in the transition from anti--phase~($i$), via traveling waves~($ii$--$iv$),
to in--phase oscillation~($v$).
In the extreme cases of $w_{13}=-1.1$ and $w_{13}=-0.92$ we can
guess once again that the firing rate is shaped by the anti--symmetric and
the symmetric transiently attracting states respectively.
In contrast to the case of lower adaption here we find that the transition
between in the in--phase and the anti--phase oscillation of the adiabatic
regime happens via the non--adiabatic regime when passing the
region of symmetry protection close to the symmetric case $w_{13}=-1$.

\subsection{Transiently attracting states in the three--neuron system}

In the bottom panels of Figs.~\ref{fig:traj_nonad}
and \ref{fig:traj_ad} we present the firing rate $y_1$ 
of the first neuron (green) and the $y_1$ component of
the corresponding target point (red bullets and line),
with the adiabatic evolution indicated by the black
arrows.

\begin{itemize}

\item For $\epsilon_\text{b}=8\cdot10^{-4}$,
      the non--adiabatic case shown in Fig.~\ref{fig:traj_nonad},
      we find that the set of target points splits into
      four branches. The jumps between the distinct
      branches are however not directly visible in the original
      trajectory, which is almost completely decoupled of
      the dynamics of the target points.

\item For $\epsilon_\text{b}=10^{-5}$, a case from the adiabatic regime
      illustrated in Fig.~\ref{fig:traj_ad}, we observe 
      on the other hand only two branches of target point
      manifolds. For extended time spans the trajectory follows closely 
      the critical manifold, jumping however periodically
      between the two distinct branches.

      One observes in Fig.~\ref{fig:traj_ad}, that the orbit needs
      a certain time, a delay, to leave a given branch of the
      target point manifold. This is a phenomenon typical for
      systems with multiple time scales \cite{kuehn2008scaling}.
\end{itemize}

\subsubsection{Stable and unstable adiabatic fixed points}

\begin{figure*}\centering
\includegraphics[width=0.75\textwidth]{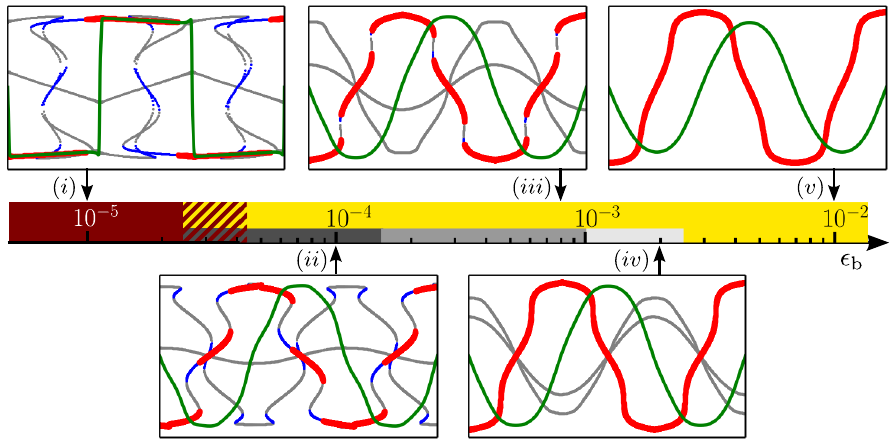}
\caption{\label{fig:phasediag_epsb} 
Horizontal cut through the phase diagram shown in Fig.~\ref{fig:phaseshift} 
at the $w_{13}=-1$ value.
The color of the bar encodes the phase shift $\varDelta_{13}$
as a function of adaption rate $\epsilon_\mathrm{b}$
(brown:$\varDelta_{13}=1/2$, yellow:$\varDelta_{13}=1/3$).
There are five phases to be distinguished by the stability and the shape of the AFP.
Each phase is represented by a characteristic trajectory--AFP--plot in $y_1$
(cf. Fig.~\ref{fig:traj_nonad} and \ref{fig:traj_ad} bottom row), the range of 
occurrence for the middle three is indicated by the gray bars within the yellow bar. 
Green/red/blue/gray in the plots denote the trajectory/target points/other 
stable AFP/unstable AFP respectively.
For relatively large values of the adaption rate~$\epsilon_\mathrm{b}$ ($v$)
there are only stable AFP, acting as the target points.
Decreasing $\epsilon_\mathrm{b}$ saddle AFP and non--target stable AFP occur
($i$--$iv$).
Hysteresis in the transition is observed in the striped area.
}
\end{figure*}

In Fig.~\ref{fig:phasediag_epsb}, which illustrates the dynamics
along the horizontal cut through the phase diagram for $w_{13}=-1$,
the adiabatic fixed points for
five different values of the adaption rate are shown. Here we have 
included all adiabatic fixed points, including the unstable ones,
which we have found by minimizing Eq.~(\ref{eq:q_flow}).
Each of the five insets shows a sketch of the trajectory (green), 
target points (red), other stable AFP (blue) and saddle AFP (gray) 
in the activity $y_1$ of the first neuron (cf. bottom panels of 
Figs.~\ref{fig:traj_nonad}, \ref{fig:traj_ad}). Each of these insets serves as an
example of the five different phases that can be distinguished by the 
shape and stability of the AFP.
The whole non--adiabatic regime with phase shift $\varDelta_{13}=1/3$ is marked by the
yellow bar, while the adiabatic regime with $\varDelta_{13}=1/2$ is indicated
by a brown bar. The range of occurrence
in the adaption rate $\epsilon_\text{b}$ of the four non--adiabatic phases is
indicated by the gray bars included in the yellow bar.

Starting at relatively high adaption $\epsilon_\text{b}=10^{-2}$
in Fig.~\ref{fig:phasediag_epsb} panel~($v$) we find only stable 
AFP of which all act as target points and the attractor metadynamics 
therefore is continuous. Decreasing the adaption rate 
$\epsilon_\text{b}$ (cf. panels ($iv$) to ($i$)) we observe the 
occurrence of saddle type AFP and stable AFP which are no targets 
to the trajectory.

In panels ($i$--$iii$) of Fig.~\ref{fig:phasediag_epsb}
one can find saddle--node bifurcations, in which stable (blue) and saddle type (gray)
AFP merge. The saddle--node bifurcation points originate from a
cusp bifurcation \cite{guckenheimer2013nonlinear} on the critical manifold,
with the slow variables as bifurcation parameters.
The adaption rate~$\epsilon_\text{b}$ determines at which point the trajectory
crosses the cusp bifurcation. Thus one either finds a continuous manifold of
target points or jumps.

This change in the structure of the corresponding target points with the adaption
rate also affects the attractor metadynamics, which becomes discontinuous as well.

\begin{itemize}
\item For all adaption rates $\epsilon_\text{b}>6\cdot10^{-5}$ we find
      stable traveling waves solutions with a relative phase difference 
      $\varDelta_{13}=1/3$, for which the trajectory is only marginally
      influenced by the transiently attracting states. This is
      the non--adiabatic regime (yellow bar).

\item For lower adaption rates $\epsilon_\text{b}<6\cdot10^{-5}$
      we find that both the trajectory and the attractor metadynamics 
      perform flip--flop oscillation with a relative phase difference 
      $\varDelta_{13}=1/2$. Here, in the adiabatic regime (brown bar),
      the motion follows strictly the metadynamics of the target points
      with some delay at the jumps, as mentioned earlier.
\end{itemize}

We emphasize that the transition between the adiabatic and non--adiabatic 
regime occurs at a remarkably low value of the adaption rate 
$\epsilon_\text{b}\approx6\cdot10^{-5}$. The traveling wave
solution being stable for $\epsilon_\text{b}>6\cdot10^{-5}$ is hence
strongly protected by the C3 symmetry of the system.
Off symmetry,
i.\,e.\ for $w_{13}\ne -1$, the transition between the adiabatic
and the non--adiabatic regime may become continuous (compare
Fig.~\ref{fig:phaseshift}), and shifts to higher
values in $\epsilon_\text{b}$.
Along the symmetric axis $w_{13}=-1$ we find a small region in the adaption
rate~$\epsilon_\text{b}$ close to the transition, as marked by stripes in
Fig.~\ref{fig:phasediag_epsb}, where attractors from both
the adiabatic and the non--adiabatic regime coexist.
Starting close to an attractor from the non--adiabatic regime we can trace
the non--adiabatic regime decreasing the adaption rate~$\epsilon_\text{b}$
in small steps. This procedure succeeds up to a certain value of $\epsilon_\text{b}$,
depending on the precision of the computation and the step size in $\epsilon_\text{b}$,
before ending up in an attractor of the adiabatic regime. Vice versa we can start
from the adiabatic regime and trace it increasing $\epsilon_\text{b}$.
This means we can observe hysteresis in the bistable transition region between
the regimes.

\subsubsection{Evolution of the average distance between
            trajectory and target points}

\begin{figure}\centering
\includegraphics[width=0.45\textwidth]{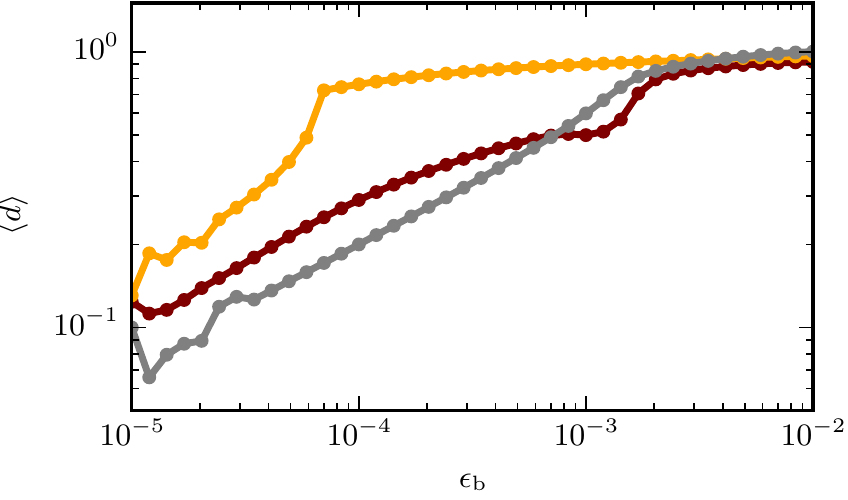}
\caption{ \label{fig:disttoAFP}
Average distance~$\left<d\right>$ of the trajectory 
to the corresponding targets, as defined by Eq.~(\ref{eq:dist}).
Shown are horizontal cuts through the phase diagram,
Fig.~\ref{fig:phaseshift}, along $w_{13}=-1.1$ (gray), 
$-1$ (yellow) and $-0.92$ (brown). The regime is non--adiabatic, 
when the distance becomes independent of 
$\epsilon_\mathrm{b}$, as it happens for larger values of the
adaption rate. The system is adiabatic, conversely, when the
distance scales with $\epsilon_\mathrm{b}$, which happens for
the symmetric case~$w_{13}=-1$ only for low adaption rates.  
}
\end{figure}

Fig.~\ref{fig:disttoAFP} shows the average distance 
$\langle d\rangle$, as defined by Eq.~(\ref{eq:dist}), 
of the trajectory to the target points. Shown
are horizontal cuts through the phase diagram
Fig.~\ref{fig:phasediag_epsb}, along 
$w_{13}=-1.1$, $-1$ and $-0.92$. 
In the absence of $C3$ symmetry, the separation of
time scales between fast and slow variables starts
to fail for larger values of the adaption rate, 
$\epsilon_\text{b}>10^{-3}$, with the distance 
$\langle d\rangle\approx1$ becoming independent of
$\epsilon_\text{b}$.

An adiabatic regime is always observed for 
low adaption rates $\epsilon_\text{b}$. It is
however clearly evident from Fig.~\ref{fig:disttoAFP}
that the adiabatic regime is pushed down for
the symmetric case $w_{13}=-1$, by more than one
order of magnitude. States protected by symmetry
operations, in our case the traveling wave solution
illustrated in Fig.~\ref{fig:traj_nonad},
can be exceedingly stable.

The discontinuous transition between the two regimes in the
symmetric case $w_{13}=-1$ is confirmed by both Fig.~\ref{fig:phaseshift}
and Fig.~\ref{fig:disttoAFP}.
There is a transition region in $\epsilon_\text{b}$, where two arbitrarily 
close initial conditions can show fundamentally different dynamics being in the 
adiabatic and the non--adiabatic regime, respectively
(cf. Fig.~\ref{fig:phasediag_epsb} and the related discussion).
On the two sides of this transition region, only one of 
these two different kinds of dynamics exists. Lacking any intermediate kind of 
dynamics, this abrupt transition is remarkable since the shape of the AFP 
manifold does not depend on the adaption rate $\epsilon_\text{b}$ at all.
We thus stress that the distance to the reference manifold 
of target points, which is unique for given parameter values and initial 
conditions, reveals a qualitative change in the dynamics.

\subsubsection{Statistics of the distance between trajectory
            and target points}

For the same parameters as used in Figs.~\ref{fig:traj_nonad} 
and \ref{fig:traj_ad} we present in Fig.~\ref{fig:disttoAFP_cummulative}
the cumulative distribution function $P(d)$ of the distance 
$d$, as defined by Eq.~(\ref{eq:cumdist}), between the trajectory 
and the corresponding target points.
First the relative density $\rho(d)$ of distances averaged over a given attractor
is computed numerically by sampling the measured distances to 400 bins 
on the logarithmic range $d\in[10^{-5},2]$. To obtain the cumulative
distribution of distances the density is then integrated
as defined by Eq.~(\ref{eq:cumdist}).

For the non--adiabatic case (magenta) we find that 
there are only contributions in a narrow region $d\in~[0.7,1.0]$ 
close to unity, resulting in turn from small residual variations 
of distance. For the adiabatic case (cyan) we observe
substantial contributions both close to unity, reflecting the 
jumps between different target branches, and extended contributions 
from small distances $d<10^{-2}$, which result from the evolution 
close to the AFP manifold. The difference between the
adiabatic and the non--adiabatic regime shows up prominently in 
the statistics of the distances between trajectory and target points.

\subsubsection{Target points corresponding to chaotic motion}

\begin{figure*}[tbh]\centering
\includegraphics[width=0.99\textwidth]{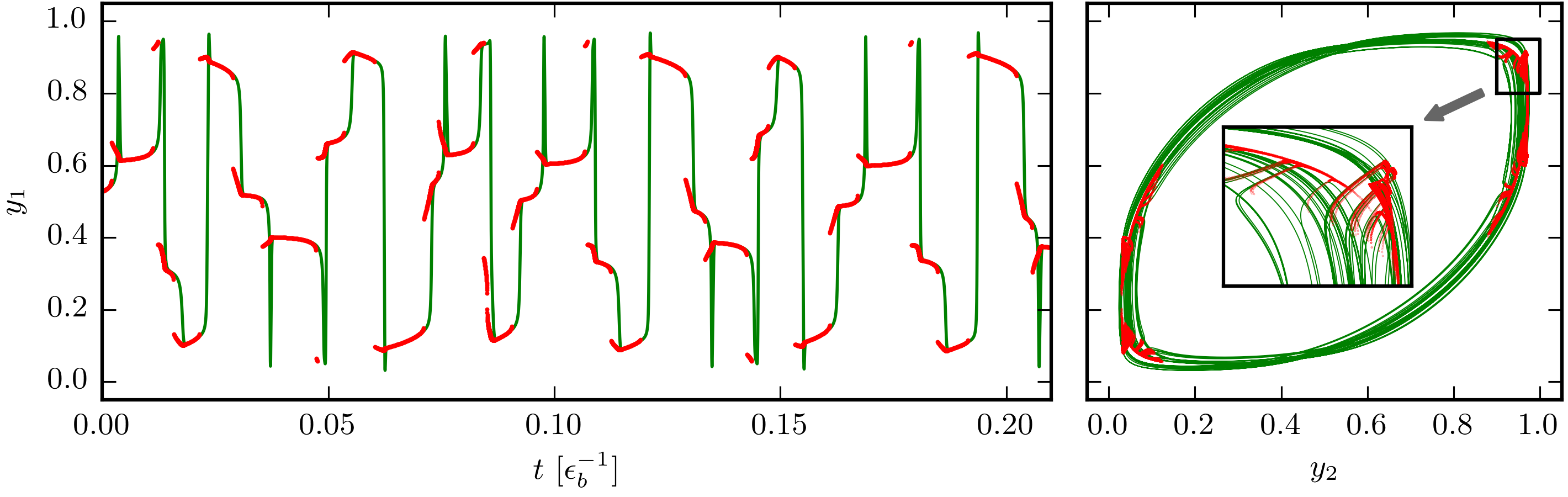}
\caption{\label{fig:trajchaos} 
Chaotic attractor (green) of 
the three--neuron system for $\epsilon_\text{b}=10^{-5}$, 
$w_{13}=-0.9709$
(indicated by the red framed triangle in Fig.~\ref{fig:phaseshift})
with the corresponding target points (red). Shown are the
firing rate $y_1$ of the first neuron over time (\textit{left})
and the projection of the orbit to the $y_1-y_2$ plane
(\textit{right}). The activity of the neuron is attracted 
towards fragments of the target point manifold, performing 
smaller and larger jumps between these fragments. The inset 
in the right--hand panel corresponds to a zoom--in to a region 
of target points, which on visual inspection seem to form a 
fractal structure.
}
\end{figure*}

For a small region in the adiabatic regime 
$\epsilon_\text{b}<3\cdot10^{-4}$ and near the symmetric 
case, $w_{13}\in[-1,-0.93]$, we find patches in the parameter 
space that exhibit chaotic motion. An example is shown
in Fig.~\ref{fig:trajchaos} for $\epsilon_\text{b}=10^{-5}$
and $w_{13}=-0.9709$. The trajectory and the corresponding
target points are shown both as a time series (left panel)
and projected to the $y_2-y_1$ plane in phase space.
Applying a recently developed $0-1$ test for chaos, based on 
the cross--distance scaling of initially nearby 
trajectories~\cite{wernecke2017test}, we find that 
this attractor is indeed chaotic.
 
Except for some overshooting, the trajectory shows a
similar behavior to the adiabatic motion presented 
in Fig.~\ref{fig:traj_ad}, approaching the target point 
manifold and staying then close to it as long as it remains
stable.

The manifold of target points has a highly non--trivial 
structure for the chaotic motion, in contrast to the 
piece--wise smooth and periodic structure observed for the case of 
limit cycle attractors, as observed e.\,g.\ in
Fig.~\ref{fig:phasediag_epsb}. A visual inspection indicates (see
the inset in the right--hand panel of Fig.~\ref{fig:trajchaos})
that the phase space projection of the target points
forms a fractal structure. We did not attempt to
directly compute the fractal dimension of the manifold
of target points shown in Fig.~\ref{fig:trajchaos}, 
as this is computationally highly demanding. 
Chaotic behavior, on the other hand, is typically 
linked to the presence of fractals in the dynamical 
behavior \cite{tel2006chaotic}.

\begin{figure}[t]\centering
\includegraphics[width=0.45\textwidth]{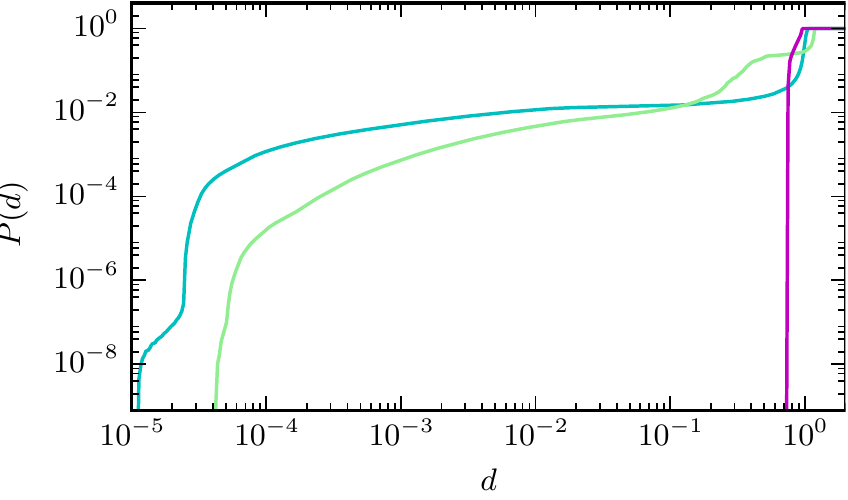}
\caption{ \label{fig:disttoAFP_cummulative}
The cumulative distribution function $P(d)$ of the distance $d$ 
between trajectory and the corresponding target point,
as defined by Eq.~(\ref{eq:cumdist}), for the adiabatic case 
($\epsilon_\text{b}=10^{-5},w_{13}=-1$, cyan), the non--adiabatic 
case ($\epsilon_\text{b}=8\cdot10^{-4},w_{13}=-1$, magenta),
and for the chaotic attractor ($\epsilon_\text{b}=~10^{-5},w_{13}=-0.9709$, 
light green; compare Fig.~\ref{fig:trajchaos}). 
For the non--adiabatic case there are only contributions close to unity
($\langle d\rangle=0.89$). Contributions from significantly 
smaller distances are observable in addition both for the adiabatic 
case ($\langle d\rangle=0.16$) and for the chaotic attractor 
($\langle d\rangle=0.25$). The latter one has contributions also
for intermediate distances $d$.
}
\end{figure}

Speaking in terms of transiently attracting states this means that in 
case of chaotic motion we cannot describe the motion as 
switching between two transiently attracting states.
But rather we find that there must be infinitely many branches of
transiently attracting states forming a fractal set.
One may nevertheless cluster these fractal sets into two broad classes,
corresponding to small and large $y_2$ respectively.
A possible interpretation of that behavior includes a chaotic motion
of the overall system driving the dynamics along the critical manifold,
which itself does not have a fractal structure.
Thus the irregular motion near the critical manifold results in a fractal
set of target points.
We emphasize that
the target points themselves are always stable fixed points of the fast
subsystem, for any values of the slow variables.

In order to compare the chaotic motion more precisely to the 
adiabatic and the non--adiabatic regimes, we have included in 
Fig.~\ref{fig:disttoAFP_cummulative} the cumulative distribution function
of distances (light green line) for the chaotic attractor. Besides 
the contribution close to unity due to the large distance jumps 
between different AFP branches, it reveals contributions at small 
distances $d<10^{-2}$, which are significant also for the adiabatic 
case (cf.~to cyan line). In this respect the chaotic attractor 
is close to the adiabatic regime. We find however additional
medium size contributions $d\approx0.3$ that result from smaller 
jumps within the fractal set of target branches.

The chaotic dynamics thus goes through an infinite series of transient states,
trying to stay close to the critical manifold (itself having a simple,
not fractal geometry), which is also confirmed
by the distribution of distances to the target points.
We do not observe chaotic exploration of the phase space, i.\,e.\ long detours
leading away from the target points as one could imagine for chaotic dynamics.
Due to the relatively small adaption rate the chaotic dynamics is guided along
different branches of the target manifold --- similar to the regular
adiabatic motion presented in Fig.~\ref{fig:traj_ad}.

Keeping on the other hand in mind that the inhibitory weight $w_{13}=-0.9709$ is
only slightly off the symmetric case, it is to be expected that the shape of
the critical manifold only changes smoothly in $w_{13}$.
Therefore we would expect the same qualitative properties for the dynamics,
i.\,e.\ the trajectory moving mostly close to branches of the target point
manifold, as in the symmetric case presented in Fig.~\ref{fig:traj_ad}.
But the origin of the chaotic dynamics is thus not obvious.


\section{Discussion and Conclusions}

We have proposed here that the study of adiabatic fixed points 
and transiently attracting states, which are sets of target points,
are useful when trying to understand complex slow--fast 
systems. As these can be realized by adding an additional slow
component to an attractor network, the target points have a
well defined physical function representing,
e.\,g., as cognitive states in a neural network.
Both the location in phase space and the topology
of the manifold of target points can be easily evaluated 
without explicit knowledge of the full set of fixed points
in the fast subsystem.
The mapping of a given trajectory onto the corresponding set of
target points is, by definition, unique and does not depend on
the time evolution on the slow time scale.
Thus target points qualify as a unique reference manifold for the
overall dynamics.
Transitions between smooth subsections of the manifold
of target points correspond generically to transitions
between well defined biological, physical or cognitive
states.

We have shown, analyzing a three--site network of adapting 
rate--encoding neurons, that system symmetries may
stabilize peculiar solutions which effectively decouple
from the dynamics of target points. This decoupling
shows up also in the statistics of the Euclidean
distance between the trajectory and the respective
target points. The distance distribution can be
used furthermore to classify states in terms of 
the relevance of the reference manifold.

We applied our analysis both to limit cycle and
to chaotic dynamics. The detailed analysis of the latter phenomenon
is a possible subject of future
work. We find, somewhat surprisingly,
that a fractal set of target points may guide 
chaotic behavior.

The above--described phenomenon is not to be confused with the case when the
attractor of the fast subsystem is more complicated than a fixed point, e.\,g.\ 
limit cycles or chaotic attractors.
In future studies the above described method for distinguishing dynamical
regimes by means of target (fixed) points could be generalized
to arbitrary target attractors.
The key implication then will be to generalize the distance measurement between
a trajectory and the corresponding attractor.

Our approach assumes a 
separation of time scales to be present, breaking
down once the time scales for the slow and for the
fast subsystem approach each other. The ability
to treat the system analytically is however not
needed.

We believe, in conclusion, that the study of target 
points could provide additional insights especially for
slow--fast dynamical systems with a relatively high number 
of degrees of freedom. These points can serve as a unique reference
to investigate the overall dynamics of a system and are easily computable.
It would be interesting to follow
the possibility to extend the here proposed approach
to non--autonomous systems, such as the modulated
neural networks processing cognitive stimuli.

%
\begin{acknowledgements}
The authors acknowledge the financial support from the
German research foundation~(DFG).
H.\,W. acknowledges the support from
Stiftung Polytechnische Gesellschaft Frankfurt~am~Main.
\end{acknowledgements}

\appendix

\section{Computing adiabatic fixed points and target points}
\label{sec:app_code}
For an alternative view we present in Table~\ref{tab:computing}
a short pseudo--code description that explains, how to effectively
compute AFP and target points. If more complex 
critical manifolds would be present, e.\,g.\ a limit cycle or a strange
attractor, one could generalize the concept~(\ref{eq:xT0}) to include
these.

\begin{table}[bt]
\caption{\label{tab:computing}Short pseudo--code description how to compute the
adiabatic fixed points (AFP) and target points.}
\renewcommand{\labelenumi}{\arabic{enumi}.}
\renewcommand{\labelenumii}{(\alph{enumii})}
\renewcommand{\labelenumiii}{(\roman{enumiii})}
\setlist[enumerate]{itemsep=0mm, leftmargin=10pt, topsep=2pt}
\noindent
\fbox{
\parbox[t][][t]{0.95\linewidth}{
\textbf{Computing AFP}
\begin{enumerate}
 \item Compute $\{\mathbf{x}(t_k),\mathbf{b}(t_k)\}_{0\leq k\leq L}$ as a solution
 of the ODE system (\ref{eq:dot_b}), (\ref{eq:dot_x}) at the time steps
 $t_k\in\{t_0,t_1,\ldots,t_L\}$\\for any $\epsilon_\text{b}>0$.
 \item For every time step $t_k$ do:
 \begin{enumerate}
  \item Set $\mathbf{b}=\mathbf{b}(t_k)$ in Eq.~(\ref{eq:dot_x}).
  \item Repeat until you found all AFP:
  \begin{enumerate}
   \item Minimize the kinetic energy~$q_x$ (\ref{eq:q_flow}) starting from random $\mathbf{x}$.
   \item Check: If the kinetic energy at the local minimum is of the order of numerical
   accuracy (but at most $q_x\approx10^{-12}$), then you found an AFP;\\
   else reject the point.
  \end{enumerate}
 \end{enumerate}
\end{enumerate} 
} 
} 
\,\\[0.3\baselineskip]
\fbox{
\parbox[t][][t]{0.95\linewidth}{
\textbf{Computing target points}
 \begin{enumerate}
 \item Compute $\{\mathbf{x}(t_k),\mathbf{b}(t_k)\}_{0\leq k\leq L}$ as a solution
 of the ODE system (\ref{eq:dot_x}), (\ref{eq:dot_b}) at the time steps
 $t_k\in\{t_0,t_1,\ldots,t_L\}$\\for any $\epsilon_\text{b}>0$.
 \item For every time step $t_k$ do:
 \begin{enumerate}
  \item Solve the ODE (\ref{eq:dot_x}) of the fast subsystem starting
  from $\{\mathbf{x}(t_k),\mathbf{b}(t_k)\}$, a point on the trajectory.
  \item Check: If the solution converged $\mathbf{\dot{x}}\to0$, you found the corresponding
  target point.
 \end{enumerate}
\end{enumerate}
} 
} 
\end{table}

\section{Fixpoints of the three--neuron system\label{sec:app_fp}}

The six dimensional three--neuron system Eqs.~(\ref{eq:dot_x}),\,(\ref{eq:dot_b})
has the only fixed point
\begin{equation}
x_2=1,\quad
x_1=x_3 = (1+w_{13})/2, \quad
b_i=x_i~.
\label{eq:fixed point}
\end{equation}
From that it follows that the neurons are half--active $y_i=1/2$
in the fixed point.
This fixed point is generally stable for large values
of $\epsilon_\text{b}$, undergoing a supercritical Hopf
bifurcation when $\epsilon_\text{b}$ becomes smaller. This
occurs, for $a=6$ and $w_{13}=-1$, at
$\epsilon_\text{b}=1/36\approx0.0278$.

\section{Traveling waves}\label{subsec:travelingwaves}\label{sec:app_tw}

One can prove that in the case of symmetric coupling $w_{13}=-1$ a traveling 
waves ansatz solves the system. This solution shows a relative phase shift of 
exactly $\varDelta_{13}=\frac{\tau}{3}$ between the first and the third neuron, 
where $\tau$ is the period of the solution.
We make an ansatz for the solution
\begin{equation}
\begin{array}{rlrlrl}
x_1(t)&=x(t-\theta),&x_2(t)&=1+x(t),&x_3(t)&=x(t+\theta)\\
b_1(t)&=b(t-\theta),&b_2(t)&=1+b(t),&b_3(t)&=b(t+\theta)
\label{eq:shift1}
\end{array}
\end{equation}

with two periodic functions $x(t)$, $b(t)$ of period $\tau$ and an arbitrary shift 
$\theta$.
The corresponding firing rates of the neurons therefore are given by
\begin{equation}
\begin{array}{rrr}
y_1(t)=y(t-\theta),&y_2(t)=y(t),&y_3(t)=y(t+\theta)
\end{array}\label{eq:shift2}\;,
\end{equation}
where the notation for the periodic function $y(t)=y(x(t),a,b(t))$ is used.
From this we get to the resulting equations of motion for the membrane potential
\begin{eqnarray}
\dot{x}(t)&=&-x(t)+\,y(t+\theta)-\,y(t+2\theta)\notag\\
\dot{x}(t)&=&-x(t)+\,y(t-\theta)+\,y(t+\theta)-1\label{eq:shift3}\\
\dot{x}(t)&=&-x(t)+\,y(t-\theta)-\,y(t-2\theta)\notag\;.
\end{eqnarray}
This leads to the condition $y(t)=1-y(t\pm3\theta)$ that has to hold when 
solving Eq.~(\ref{eq:shift3}).
Assuming further that $x(t+\frac{\tau}{2})=-x(t)$ and 
$b(t+\frac{\tau}{2})=-b(t)$, one can solve Eq.~(\ref{eq:shift3}) for the 
constant phase shift $\theta=\frac{\tau}{6}$.
Therefore the phase shift $\varDelta_{13}$ between the first and the third
neurons is exactly $\varDelta_{13}=2\theta/\tau=1/3$.



%

\end{document}